\documentclass[oneside]{article}
\usepackage{amsfonts}

\newtheorem{theorem}{Theorem}{}
{}
{}
{}
{}
{}

\newcommand{\ra}{\rightarrow}

\title{{\bf Correct by construction}}
\author{M.H. van Emden\\
        {\small Technical Report DCS-361-IR}\\[-1mm]
        {\small Department of Computer Science}\\[-1mm]
        {\small University of Victoria}
       }
\date{}

\begin{document}
\maketitle
\date{}

\begin{abstract}
Matrix code allows one to discover algorithms and to render
them in code that is both compilable and is correct by construction.
In this way the difficulty of verifying existing code is avoided.
The method is especially important for logically dense code
and when precision programming is called for.
The paper explains both these concepts.
Logically dense code is explained by means of the partition stage
of the Quicksort algorithm.
Precision programming is explained by means of fast exponentiation.
\end{abstract}

\section{{\large Introduction}}\label{sec:intro}
In 1969 E.W. Dijkstra introduced \cite{dijkstra249}
Structured Programming.
At the time the method was revolutionary:
it advocated the elimination of the goto statement
by the exclusive use of conditional, alternative,
and repetitive clauses.
It took only a few years for structured programming
to evolve from heresy to orthodoxy.
But by then Dijkstra had arrived at the conviction
that the looming ``software crisis'' could only be averted
by formally verifying code.
At the time Hoare's method \cite{hoareAxiomatic}
was the only one available for structured programs.
It was found difficult to use in practice.
Dijkstra maintained his conviction that code needed to
be verified.
He defended it in his 1971 paper
``Concern for correctness as guiding principle in program
composition'' \cite{dijkstra71}.
For my purpose the crucial passage is
\begin{quotation}
{\sl
When correctness
concerns come as an afterthought and correctness proofs have to
be given once the program is already completed, the programmer
can indeed expect severe troubles. If, however, he adheres to the
discipline to produce the correctness proofs as he programs along,
he will produce program and proof with less effort than programming
alone would have taken.
}
\end{quotation}
Let us give the name ``correctness-oriented programming'' to
this hypothetical method of producing
{\sl the correctness proofs as he programs along}.

Within imperative programming the only possibility of
correctness-oriented programming is programming with verification
conditions \cite{back09,reynolds78,vanemden79,vanemden14}.
In \cite{vanemden14} programming
with verification conditions is furthest evolved and receives
the name ``Matrix Code''.
Yet even then it needed further improvement to make it
practically applicable.
This was done in \cite{vanemden18}.
The present paper presents two situations where programming with
verification conditions is especially useful:
precision programming and logically dense code.
These concepts are explained in Section~\ref{sec:motivation}.
In Section~\ref{sec:MC} we give
a self-contained exposition of Matrix Code.

\section{{\large Matrix Code}}\label{sec:MC}

Before Dijkstra can make good on his promise,
we will have to find a way out of a chicken-and-egg situation:
a correctness proof has to be developed for
a program that does not yet exist.
Such a way was first proposed,
according to R-J. Back \cite{back09},
by J. Reynolds \cite{reynolds78}
and by M.H. van Emden \cite{vanemden79}.
Van Emden's method, ``programming with verification
conditions'',
was developed into the Matrix Code of 2014 \cite{vanemden14}.
In this paper we follow the practical implementation of
Matrix Code shown in \cite{vanemden18}.

\subsection{{\normalsize Floyd's verification method}}
In 1967 R.W. Floyd proposed a method for verifying
programs \cite{floyd67}.
Here verification consists of a mathematical proof
that the final state has a certain property
provided that the initial state satisfies certain conditions.
The program is in the form of a flowchart
consisting of tests and statements
that read from and write to a set of shared variables
that together comprise the state.
Without loss of generality we can assume that
there is one node, the start node, without incoming arc.
Likewise 
there is one node, the halt node, without outgoing arc.
Some of the nodes are provided with labels.
This has to be done in such a way that there is no pair of
labels between which there is an infinite path
without a label.
The start node gets label S; the halt node gets label H.

Floyd's method associates each label with an assertion.
An assertion asserts that a certain relation holds between
some of the variables.
Assertions are expressed in the form of a formula
of first-order predicate logic
in which the free variables have the same names
as some of the flowchart's variables.
Floyd's method assumes an interpretation for the function
symbols and predicate symbols occurring in the assertions
associated with the labels of the flowchart.

There is a ``verification condition'' associated with every
pair $\langle P,Q\rangle$ of labels connected by a path
that does not contain any label.
The verification condition has the form $\{p\}S\{q\}$
where $p$ and $q$ are the assertions associated with $P$
and $Q$ and where $S$ is the binary relation between states
resulting from the execution of the code in the path from
$P$ to $Q$.
The verification condition $\{p\}S\{q\}$ is true
if and only if the truth of $p$ in a state $s$
and the execution of $S$ starting in $s$ results in
state $t$ implies that $q$ is true in $t$.

Floyd's verification method is embodied in the following
theorem.
\begin{theorem}
Given a flowchart $F$ with start label $S$ and halt label $H$.
If the verification conditions of $F$ are true,
then $\{s\}F\{h\}$ is true, where $s$ and $h$ are the assertions
associated with $S$ and $H$.
\end{theorem}
A verification condition can be regarded as a fragment of
executable code.
Compare $\{p\}S\{q\}$ with
{\tt P: execute "S"; goto Q},
where {\tt P} labels assertion $p$,
where {\tt Q} labels assertion $q$, and
{\tt S} is the code connecting {\tt P} to {\tt Q}.
The value of this observation lies in the fact that
{\tt S} need not come from a conventional flowchart.
It can be anything that has a binary relation among states
as meaning.
In particular, this relation can be a subset of the identity
relation.
For example \verb+{p} x>0 {p & x>0}+
is defined and is true for any {\tt p}.

What is true for a single verification condition also
holds true for sets of verification conditions.
Such a set can start small. It can then point in the direction
of a missing verification condition,
which can then be added.
This a way of following up on Dijkstra's hint
``produce the proof as he programs along''.

\subsection{{\normalsize Matrix Code}}
A set of verification conditions with generic element
\verb+{P}C{Q}+
can be read as the specification of a matrix with columns
and rows indexed by the labels.
{\tt C} is then the matrix element in column {\tt P}
and row {\tt Q}.
Ergo, a set of verification conditions is a program in
Matrix Code.
Let us now investigate Matrix Code as a programming
language.

\paragraph{Syntax of Matrix Code}
A {\sl program} consists of a signature, a set of assertion
declarations, and a set of triples of the form
\verb+{P}C{Q}+,
where {\tt P} and {\tt Q} are labels and {\tt C} is a command.
The signature consists of a set of function symbols,
a set of predicate symbols, and a set $V$ of variables.
The assertion declarations associate an assertion with
every label.
An assertion is a formula.
A command is a formula, an expression of the form
$v := \tau$, where $v\in V$ and $\tau$ is a term,
or an expression of the form $C;D$, where $C$ and $D$
are commands.

Formulas, terms, and variables are defined as in first-order
predicate logic with respect to the signature of the program. 

\paragraph{Semantics of Matrix Code}
The meaning of a program is truth or falsity with respect
to an interpretation $I$ for the signature of the program.
The meaning of an assertion is a set of states.
The meaning of a command is a binary relation over the set
of states.

The set of states is the set of tuples of type $V\ra D$,
where $D$ is the universe of discourse of $I$.

The meaning of an assertion $F$ in $I$ is
$\{s\in (V\ra D) \mid
     F\mbox{ is true in } I\mbox{ in state } s
 \}
$.
If a command is a formula $F$, then its meaning in $I$ is
$\{\langle s,s\rangle \mid s\in (V\ra D) \wedge
     F\mbox{ is true in } I\mbox{ in state } s
 \}
$.
If a command is $v := \tau$, then its meaning is
$\{\langle s,s^v_\tau\rangle \mid s\in (V\ra D)\}$
where $s^v_\tau$ is defined by $s^v_\tau(x) = s(x)$
if $x\in V$ is not $v$ and
$s^v_\tau(v)$ equals the value of the term
$\tau$ in $I$ in the state $s$.
If a command is $C;D$, then its meaning is the relational
product of the meanings of $C$ and of $D$.

``$F$ is true in $I$ in state $s$'' and
``the value of the term $\tau$ in $I$ in the state $s$''
are defined as in logic.

\section{{\large Motivation for programming in Matrix Code}}
\label{sec:motivation}

\subsection{{\normalsize Dense logic}}

Some code is harder to get right than other code.
It would help to know more about
how, and where errors occur.
Documentation on this is hard to come by.
It is in the nature of the situation:
as long as the bug is still alive, one is frantically chasing it.
As soon as it is found,
it is seen to be an egregious failing of the most
elementary intelligence, too embarrassing to confess.

This makes Jon Bentley's confession \cite{bentley86} (chapter
{\sl Sorting}) valuable:
\begin{quote}
$\ldots$ I once spent the better part of two days
chasing down a bug hiding in a short partitioning loop.
\end{quote}
I suspect Bentley is not the only one,
but I know of no similar published confessions.
Here is my experience with Quicksort.
In 1968 it was many times that I ran up and down the stairs
between my desk
on the fourth floor and the computer room on the ground
floor in chaotic attempts at debugging a partition function.
Accepting a version of CACM Algorithm 402 as correct was based
on checking whether it had correctly sorted an array of length
thirty-six, the longest that would
fit on a single line of line-printer paper
so that it could be checked at a glance.

In this paper we examine the Quicksort algorithm.
In developing code for two versions of the partitioning function
Dijkstra's
``{\sl he will produce program and proof with less effort
than programming alone would have taken}'' will turn out
to be not an empty promise.

\subsection{{\normalsize Desire for precision programming}}

Structured programming replaces a versatile precision instrument
for control by a few coarse-grained primitives.
Initially three constructs were allowed that were borrowed from
Algol 60.
Later, Dijkstra replaced these by his language of guarded commands.
In either form, structured programming is inimical to precision
in control.

This claim is of course vacuous until ``precision programming''
has been demonstrated by an example.
The example I use is:
\paragraph{}
given a positive number $X$ and a nonnegative integer $Y$,
set $z$ equal to $X^Y$ by means of repeated multiplication
\paragraph{}
In \cite{dijkstraFeijen} (page 60)
the authors arrive at the following program.
\begin{verbatim}
|[ x,y: int; z:=1; x:=X; y:=Y
 ; do y mod 2 != 0 -> z := x*z; y := y-1
    | y != 0 & y mod 2 = 0 -> x := x*x; y := y/2
   od
]|
\end{verbatim}
There can be two obstacles to understanding this code:
lack of familiarity with guarded commands and
its contorted structure. The latter is due to its derivation.
Without the desire to derive it,
one would write:
\begin{verbatim}
|[ x,y: int; z:=1; x:=X; y:=Y
 ; do y != 0 ->
      if y mod 2 != 0 -> z := x*z; y := y-1
       | y mod 2  = 0 -> x := x*x; y := y/2
      fi
   od
]|
\end{verbatim}
For those not familiar with guarded commands,
this transcribes to C as follows.
\begin{verbatim}
int x,y; z = 1; x = X; y = Y;
while (y != 0) {
  if ((y%2) != 0) { z = x*z; y = y-1; continue; }
  if ((y%2) == 0) { x = x*x; y = y/2; continue; }
}
\end{verbatim}
The authors note that the computation {\tt x*z} is not necessary
the first time because {\tt z} is 1.
Also, after {\tt y := y-1} that variable is even,
and this fact is not used.
They continue with ``All such attempts [at improvement]
probably make the program
text less clear, and in any case much longer.''
This is apparently such a serious obstacle
that the authors leave it the way it is.
But rather than noting that structured programming comes
with drawbacks, they continue with:
\begin{quote}
There was a time when constructing the fastest program,
irrespective of cost, was considered profoundest wisdom.
Now, fortunately, this has become obsolete.
\end{quote}
Perhaps the authors have decided that attack is the
best defence---if these flaws bother you,
then there is something wrong with {\sl you},
not with their program!
What is demonstrated here is that for this problem the
greater degree of precision needed is discouraged by the
rigid corset imposed by structured programming.

\section{{\large Egyptian multiplication: 
  an example of precision programming}}\label{sec:exPP}
We have illustrated precision programming by an example where
it is lacking.
We continue with an example where it is practised:
Stepanov and Rose \cite{stepRose} address a similar problem
without evading its complications.
At the same time they work at a higher level of generality,
noting that the trick that expedites the program just discussed
has wider applicability. They express it thus:
to compute $na$, where $n$ is a nonnegative integer and $a$ is
a number on which $+$ is defined.
They call it ``Egyptian multiplication''.
The trick is to exploit the associativity of $+$:
$na$ can be defined as
$$
a+a+\cdots+a
$$
This can be computed as
$$
a+(a+(a+(a+(a+(a+(a+a))))))
$$
requiring seven operations or as
$$
((a+a)+(a+a))+((a+a)+(a+a))
$$
requiring three operations.
More generally, in the order of $n$ compared to $\log n$ operations.

To get greater benefit of the general applicability of the idea,
Stepanov and Rose use the facilities for generic programming
in C++ to make their code applicable to monoids
(algebras where $+$ is associative).
They go one step further and note that the case $n=0$ is not
needed.
By assuming that $n>0$ the code will not need the operation's
neutral element, so that the code is no longer restricted
to monoids but becomes applicable to semigroups in general.

To lighten the load of language detail
we refrain from the use of the generic facilities of C++
and compute $na$ with positive integer $n$ and a floating-point
type for $a$ as a sort of symbolic stand-in for the semigroup
type.
Its symbolic status is underlined by the fact that addition in
the floating-point standard is not associative.

As Matrix Code, being a matrix, has no conveniently printed form,
we write sets of verification conditions,
which are only trivially different.
Below we develop several successive versions.

Initially we only have the problem statement,
so only the start label and the halt label
with associated assertions.
\begin{verbatim}
Label declarations
------------------
S: n = n0 & a = a0 & n0 > 0
H: n0*a0 has been printed
\end{verbatim}
We cannot think of any sufficiently simple code {\tt C} that
ensures \verb+{S}C{H}+,
so need to add assertions, which come with associated new labels.

In deciding what these are going to be we note
that the problem would be simplest in the case where $n$
is a power of 2:
then we only need to double $a$ a suitable number of times.
And in the likely event that $n$ is not initially a power of 2,
it is still efficient to start by removing all factors of 2
from $n$.
\begin{verbatim}
Label declarations
------------------
S: n = n0 & a = a0 & n0 > 0
A: n0*a0 = n*a & n > 0
B: A & odd(n)
H: n0*a0 has been printed

Triples
-------
{S} skip {A}
{A} even(n); a := a+a; n := n/2 {A}
{A} odd(n) {B}
{B} n = 1; print a {H}
\end{verbatim}

This is already an executable program, although it only works
for {\tt n} a power of two.
To handle the general case
we need to add another triple emanating from {\tt B}
and to introduce a variable {\tt z}
to collect part of final result;
see the additional assertion labeled with {\tt C}.

\begin{verbatim}
Label declarations
------------------
S: n = n0 & a = a0 & n0 > 0
A: n0*a0 = n*a & n >0
B: A & odd(n)
C: n0*a0 = z+n*a & n > 0
H: n0*a0 has been printed

Triples
-------
{S} skip {A}
{A} even(n); a := a+a; n := n/2 {A}
{A} odd(n) {B}
{B} n = 1; print a {H}
{B} n != 1; z := a; a := a+a; n := (n-1)/2 {C}
\end{verbatim}

The algorithm proper can start with triples
emanating from {\tt C}, which gives rise to
a new assertion {\tt D} and triples emanating from it.

\begin{verbatim}
Label declarations
------------------
S: n = n0 & a = a0 & n0 > 0
A: n0*a0 = n*a & n > 0
B: A & odd(n)
C: n0*a0 = z+n*a & n > 0
D: C & odd(n)
H: n0*a0 has been printed

Triples
-------
{S} skip {A}
{A} even(n); a := a+a; n := n/2 {A}
{A} odd(n) {B}
{B} n = 1; print a {H}
{B} n != 1; z := a; a := a+a; n := (n-1)/2 {C}
{C} odd(n) {D}
{C} even(n); a := a+a; n := n/2 {C}
{D} n = 1; print z+a {H}
{D} n > 1; z := z+a; a := a+a; n := (n-1)/2 {C}
\end{verbatim}

No more labels are needed; no triples are missing.
The items needed to reach this result became apparent
as we started from the initial version that only contained
the problem statement.

\section{{\large Liffig form of Matrix Code}}\label{sec:liffig}

The simplicity of the Matrix Code syntax causes it to allow
some unfamiliar forms---forms that defy simple transcription
to one's favourite implemented language.
For example $x>0 ; x<11$ is a legal command with a perfectly
well-defined meaning.
The same can be said of $ x := x+1; x < 10$,
which has the same meaning as $ x < 9; x := x+1$
in the interpretation with the mathematical integers as
universe of discourse.
The latter form is preferred for ease of transcription.

``Liffig'' is the name of a notation for Matrix Code programs
that differs in encouraging ease of transcription.
It imposes a helpful order on the set of triples.
It integrates the label declarations with the set of triples.
It gives priority to preferable forms of commands.
Liffig is inspired by Dijkstra's guarded commands. 
As a sample I present the Liffig form of the Egyptian
multiplication example.
See Figure~\ref{prog:egyptMult}.

\begin{figure}
\begin{center}
\begin{minipage}[t]{4in}
\hrule \vspace{2mm}
\begin{verbatim}
S: n = n0 & a = a0 & n0 > 0
  if true -> skip; goto A
  fi
A: n0*a0 = n*a & n > 0
  if  even(n) -> a := a+a; n := n/2; goto A
   |  odd(n) -> goto B
  fi
B: A & odd(n)
  if  n = 1  -> print a; goto H;
   |  n != 1 -> z := a; a := a+a; n := (n-1)/2; goto C
  fi
C: n0*a0 = z+n*a & n > 0
  if  odd(n) -> goto D
   |  even(n) -> a := a+a; n := n/2; goto C
  fi
D: C & odd(n)
  if n = 1 -> print z+a; goto H
   | n > 1 -> z := z+a; a := a+a; n := (n-1)/2; goto C
  fi
H: n0*a0 has been printed
  return
\end{verbatim}
\hrule
\end{minipage}
\end{center}
\caption{
\label{prog:egyptMult}
The program for Egyptian multiplication in Liffig form.
}
\end{figure}

The reader will recognize the {\tt if ..|.. fi} of Dijkstra's
guarded commands. Each of these is preceded by a label.
Each individual guarded command is terminated by
{\tt goto L} to indicate that the assertion labelled by {\tt L}
holds at that point.
Because of the added label before {\tt if ..|.. fi} and the
added {\tt goto}'s following I have named the language
``Liffig'', to be parsed as ``L/iffi/g''.

In the sequel I will use Liffig to develop two versions of
the partition function of Quicksort.
It has the advantage that its verification conditions
can be clearly recognized. It is easy to
transcribe to several implemented programming languages.

\section{{\large Quicksort}}\label{sec:QS}
The Quicksort algorithm is
a common example in introductory programming textbooks;
it is also important in practice.
The idea of the algorithm is that to sort an array, 
one begins by swapping elements in such a way
that the array ends up being {\sl partitioned}.
This means that the array is decomposed into two parts
in such a way that nothing in the left part is greater than
anything in the right part.
In this way the problem of sorting the entire array
is reduced to the subproblems of sorting the left part
and of sorting the right part.
These two problems are smaller and can be solved by
the same method.

The problem reduction has a simple form in a suitable programming
language. Hoare used Algol 60; a C version follows below.
\begin{verbatim}
void Quicksort(int A[], int m, int n) {
  int i, j;
  if (m<n) {
    partition(A, &i, &j, m, n);
    Quicksort(A, m, j);
    Quicksort(A, i, n);
  }
}
\end{verbatim}

Thus the only work that remains is the function
{\tt partition}.
This can be found in dozens of introductory
textbooks for programming.
It so happens that the two main variants
have been authored or co-authored by Hoare himself,
which is my reason for restricting the treatment to these.

%
\section{{\large Partitioning tactics}}

An abstract statement of the goal of partitioning is
\begin{quote}
the array is decomposed into two parts
in such a way nothing in the left part is greater than
anything in the right part.
\end{quote}
This implies that there is an item of the same type as the
array elements (so that they can be compared) such that
nothing in the left part exceeds the item nor anything
in the right part is exceeded by the item.
Such an item is called a ``pivot''.
So defined, the pivot may or may not itself be an array
element.

For the choice of the pivot
there are practical and theoretical considerations.
Practical considerations include
whether one wants to optimize for some special
property that the array to be sorted might have.
The properties that might arise in practice include
that there may be few or many equal items
and that the array may be random or nearly sorted.
We next turn to theoretical considerations.

%

Partitioning results in an array of length $n$ being
subdivided into segments of lengths $r$ and $n-r$.
Before partitioning the number of possible permutations is
$n!$---afterwards this is $r!(n-r)!$.
Information theory measures the uncertainty in not knowing
which of $N$ equally probable possibilities is actually
the case as $\log N$ bits.
Thus the reduction of uncertainty achieved by the partition is
\begin{equation}\label{eq:middle}
\log n! - \log r! - \log (n-r)! = \log \frac{n!}{r!(n-r)!}.
\end{equation}
This is maximal for $r=n-r$;
when the partition ends in the middle.
To achieve this the partition algorithm needs a clairvoyant
choice of pivot. An example of a realizable choice of pivot
causes $r$ to assume every value in $1,\ldots,n$ with equal
probability. It can be shown that this makes (\ref{eq:middle})
smaller than its maximum value by a factor of $1.386$
\cite{hoareCompJ}.

One can get a larger yield out of a partition by
choosing as pivot the median of a random sample of size $m$ out of
the $n$ array elements.
The resulting gain in Quicksort performance has, of course,
to be balanced against the cost of computing the median.
Strategies for choosing the pivot and adaptations of Quicksort
to the several properties of the sequence to be sorted
represent a vast area for research;
one could devote a whole thesis to it\footnote{
In fact, somebody did; see
``Quicksort'' by Robert Sedgewick. PhD thesis, Stanford, 1975.
}.

\section{{\large Hoare's 1961 partition}}
\label{sec:1961Part}
The first version of Quicksort was published in 1961 as
Algorithms 63 and 64 \cite{hoareAlg63,hoareAlg64}.
Curiously enough Google Scholar gives about a dozen references,
to be compared with over a thousand to the journal article
\cite{hoareCompJ},
which contains much ancillary information, but no code.
It turns out that Algorithm 63 is quite interesting;
see Figure~\ref{prog:alg63Spec}.
\begin{figure}
\begin{center}
\begin{minipage}[t]{4in}
\hrule \vspace{2mm}
\begin{verbatim}
procedure partition (A,M,N,I,J); value M,N;
  array A; integer M,N,I,J;
comment I and J are output variables, and A is the array (with
subscript bounds M:N) which is operated upon by this procedure.
Partition takes the value X of a random element of the array A,
and rearranges the values of the elements of the array in such a
way there exist integers I and J with the following properties:
      M <= J < I <= N provided M < N
      A[R] <= X for M <= R <= J
      A[R]  = X for J < R < I
      A[R] >= X for I <= R <= N
\end{verbatim}
\hrule
\end{minipage}
\end{center}
\caption{
\label{prog:alg63Spec}
Hoare's specification for the partition algorithm in
Algorithm 63.
}
\end{figure}
Multiple elements equal to the pivot {\tt X} can end up in the middle
and will not be touched by future recursive calls by Quicksort.
It would be nice if this could be said for {\sl all} elements
equal to {\tt X}. The above listing does not say.

Perhaps Algorithm 63 does have this property.
It seems a daunting task to prove this.
See the listing in Appendix~\ref{sec:origQS}.
In fact, it seems a daunting task to even prove that the algorithm
satisfies the weaker property claimed.
This is only to be expected with code not constructed
with a correctness proof in mind.
Let us now embark on a derivation of a Matrix Code program.

Here I have the benefit of a sequel to Algorithm 63 in the form
of a paper \cite{stepanov06} by Stepanov in which he shows that
the well-known problem of the Dutch National Flag \cite{dijkstra76}
is a disguise
for Hoare's Algorithm 63 {\sl as it should be done}.
The ``Flag'' problem is to arrange an array of which the elements
have at most three different elements.
We will distinguish them as red, white, and blue.
The task is to permute them so that there is no red to the right of
any white and no white to the right of any blue.
Thus, if all three occur, then in the desired state
all the reds are at the left (lower indexes),
all the blues are at the right (higher indexes),
and all the whites in between.

Stepanov noticed that red, white, and blue can be read as encodings
for less than, equal to, and greater than the pivot in Algorithm 63.
In addition to this encoding, Dijkstra goes to great lengths
to hide the connection; he speaks in terms of buckets,
patriotically coloured pebbles,
and a mini-computer to inspect contents of buckets.

Now that the secret is out, a good way to improve Algorithm 63
is to write a program for the Dutch National Flag problem
for array segments containing no items other than those I shall
call {\sl red}, {\sl white}, and {\sl blue}.
I follow the idea of Stepanov's code \cite{stepanov06} (page 209),
but will write the program in Liffig. In the course of transcription
to C the array will become an array of integers, a pivot {\tt X}
will appear, and {\sl red}, {\sl white},
and {\sl blue} will be translated to less
than, equal to, and greater than {\tt X}.

Initially the program is only a problem statement.
\begin{verbatim}
S: a[m..n] is allocated, with m <= n &
   a[m..n] contains no elements other than red, white, blue
   ...
H: a[] is a permutation of a0[] &
  there exist f, s, and t such that m<=f<=s<=t<=n+1 & 
  a[m..f-1]=red & a[f..s-1]=white & a[t..n]=blue
\end{verbatim}
The problem is to find suitable text to replace the dots.
The main item is an assertion {\tt A} 
that is easy to reach from {\tt S} and
that is easy to adjust in the direction of {\tt H}. 
\begin{verbatim}
S: a[m..n] is allocated, with m <= n &
   a[m..n] contains no elements other than red, white, blue &
   a[] = a0[]
  if true -> f := m; s := f; t:= n+1; goto A
  fi
H: a[] is a permutation of a0[] & m<=f<=t<=n+1 & 
  a[m..f-1]=red & a[f..t-1]=white & a[t..n]=blue
  if true -> *j = f-1; *i = t; return
  fi
A: s <= t &
  a[] is a permutation of a0[] & m<=f<=s<=t<=n+1 &
  a[m..f-1]=red & a[f..s-1]=white & a[t..n]=blue
\end{verbatim}
Assertion {\tt A} can be illustrated 
with an array of 12 elements, as shown below.
\begin{verbatim}
        array content  R R R W W W U U U B B B
                       |     |     |     |   |
              index    m     f     s     t   n
\end{verbatim}
The array contents,
red, white, unknown, and blue, are shown as 
{\tt R},
{\tt W},
{\tt U}, and
{\tt B}
respectively.
The indexes 
{\tt m},
{\tt f},
{\tt s},
{\tt t}, and
{\tt n}
have the names of the corresponding variables in the code.

Code has been added at {\tt S} to establish {\tt A}.
The code at {\tt H} is added to pass the result of the
computation to the call of the function of which we are 
writing the body.

At {\tt A} the task is to make the gap between {\tt s} and
{\tt t} smaller.
We first pick off the easy case {\tt s=t}, 
where there is no need to make the gap smaller.
The case {\tt s<t} is almost equally simple.

See Figure~\ref{prog:dnf}.

\begin{figure}[h]
\begin{center}
\begin{minipage}[t]{4in}
\hrule \vspace{2mm}
\begin{verbatim}
S: a[m..n] is allocated, with m <= n &
   a[m..n] contains no elements other than red, white, blue &
   a[] = a0[]
  if true -> f := m; s := f; t:= n+1; goto A
  fi
H: a[] is a permutation of a0[] & m<=f<=t<=n+1 &
  a[m..f-1]=red & a[f..t-1]=white & a[t..n]=blue
  if true -> *j = f-1; *i = t; return
  fi
A: s <= t &
  a[] is a permutation of a0[] & m<=f<=s<=t<=n+1 &
  a[m..f-1]=red & a[f..s-1]=white & a[t..n]=blue
  if s = t -> *j := f-1; *i := t; return
   | s < t -> goto B
  if
B: A & s < t
  if a[s]=red -> swap(a,f,s); f := f+1; s := s+1; goto A
   | a[s]=white -> s := s+1; goto A
   | a[s]=blue -> t := t-1; swap(a,s,t); goto A
  fi
\end{verbatim}
\hrule
\end{minipage}
\end{center}
\caption{
\label{prog:dnf}
Liffig code for the Dutch National Flag problem.
See Appendix~\ref{app:dnfQS} for the transcription to C.
}
\end{figure}

\section{{\large Foley and Hoare partition}}
\label{sec:FHpart}
I have written at length about how the first Hoare partition,
the one in Algorithm 63 \cite{hoareAlg63},
can be improved by identifying it with the problem of the Dutch
National Flag.
As a result the main idea of Quicksort was eclipsed by the
Dutch National Flag.
Many implementations of Quicksort do not follow Algorithm 63,
but are more like the version published by Foley and Hoare
\cite{foHoare71}. It can be described as follows.

After choosing the pivot, an upward sweep is performed until
an obstacle is encountered in the form of an element that is
greater than the pivot.
A downward sweep is performed until
an obstacle is encountered in the form of an element that is
smaller than the pivot.
With both sweeps stopped, the obstacles are swapped.
The two sweeps and subsequent swaps are repeated until
the unpartitioned segment vanishes. 

The upward and downward sweeps do two comparisons per array element
in Algorithm 63: 
one comparison with the pivot and one comparison
to ensure that the next index does not exceed the array bound.
This work is rewarded when the array is long,
but has relatively few different values.
When multiple values are rare, it pays to avoid checking
for exceeding the array bound by means of a ``sentinel''.
In the outline sketched above only an array element {\sl greater}
than the pivot is counted as an obstacle in the upward sweep. 
Suppose that this criterion is changed to ``not smaller'',
then the comparison with the array bound can be dispensed with.
With the analogous modification applied to the downward sweep
we get a partition in which only one comparison is needed per
array element. The cost of this speed-up is that the exchange
of obstacles is also done when these are equal (to each other
because equal to the pivot).
Using the pivot as sentinel has two advantages.
The first is that often the array to be sorted is
such that equal elements are rare.
This raises the question why to choose as pivot an
array element in the first place.
This is easy to avoid, which is another reason for choosing the
sentinel version of partitioning.

In Section~\ref{sec:1961Part} a Liffig program was developed from a
specification informally obtained from the published code
in \cite{hoareAlg63,hoareAlg64}.
I showed some intermediate steps in the development of the
Liffig program.
Now is the time to do the same on the basis of the code
published in \cite{foHoare71}, except that no intermediate
steps will be shown. 
See Figure~\ref{prog:dnf}.
\begin{figure}
\begin{center}
\begin{minipage}[t]{4in}
\hrule \vspace{2mm}
\begin{verbatim}
S: m<n & a[m..n] allocated & a = a0
   if true -> i := m; j := n; r = (a[m]+a[n])/2;
              goto A
   fi
H: a is perm. of a0 & i = j+1 & a[m..j] <= r <= a[i..n]
   if true -> return
   fi
A: a is perm. of a0 & i <= j & a[m..i-1] <= r <= a[j+1..n]
  if i = j -> goto F
   | i < j -> goto B
  fi
B: A & i<j
  if a[i] < r -> ++i; goto A;
   | a[i] >= r -> goto C
  fi
C: B & a[i] >= r & i<j i.e. blocked on left
  if a[j] > r -> --j; goto D
   | a[j] <= r -> { blocked on left and right & i<j }
     swap(a, i, j); ++i; --j; goto E
  fi
D: A & a[i] >= r & i <= j
  if i = j -> goto F
   | i < j -> goto C
  fi
E: a is perm. of a0 & a[m..i) <= r <= a(j..n]
  if i > j -> goto H
   | i <= j -> goto A
  fi
F: A & i=j
  if a[i] <= r -> ++i; goto H
   | a[j] >= r -> --j; goto H
  fi
\end{verbatim}
\hrule
\end{minipage}
\end{center}
\caption{
\label{prog:foHo}
Liffig code for the partition function of Foley and Hoare
\cite{foHoare71}.
}
\end{figure}

\section{{\large Concluding remarks}}

\paragraph{Scope of Matrix Code}
Commands are limited to assertions and to assignments in which
the right-hand side is a term of logic.
This makes Matrix Code a marvel of simplicity,
syntactically and semantically.
It can only be recommended for writing bodies of
function definitions in another language.
This other language is needed to make the Matrix Code
program into a procedure body.

\paragraph{Relation of Matrix Code to logic}
An intriguing aspect of Matrix Code, the purest form of
imperative programming, is that so much of it is defined
by first-order predicate logic, a formalism that antedates
programming languages by decades.
Syntactically, signatures, formulas (assertions and commands),
variables, terms are adopted from this logic.
Semantically, the value of a term and the truth of a formula
are defined as in Tarski's semantics for logic.
Added to logic are {\tt :=}, {\tt ;}, as well as the notion
of {\sl program} on the syntactic side.
On the semantic side, states and binary relations over them
are foreign to logic.

\paragraph{Relation of Matrix Code to logic programming}
An important advantage of Matrix Code is that it can grow
from small beginnings by accretion of verification conditions.
Even such small beginnings have computations. These cannot
give incorrect results as long as the verification conditions
are true.
But they can fail to give desired answers
as long as not all requisite verification conditions
have been added.

Logic programming has the same advantage. The programs grow by
accretion of facts or parts of
a procedure definition which have meaning by themselves
as a Horn clauses of logic and as such are true or false. 
The similarity between flowchart programs and a certain
style of logic programs was noted in \cite{clarkVE}.

\paragraph{Relation to matrix theory}
So far in this paper it was only noted that verification
conditions, being of the form $\{P\}S\{Q\}$,
can be read as a listing of items specifying a matrix with
element $S$ in column $P$ and row $Q$.
In \cite{vanemden14} it is shown that there is a deeper
justification of viewing collections of verification conditions
as matrices. For example, if $M_1$ and $M_2$ are matrix code
programs on the same vocabulary, then their matrix product is
defined and denotes a matrix code program.
The assertions associated with the labels can be regarded as
a column vector $A$.
The equation $MA = A$ can be interpreted as a correctness proof
of program $M$ with respect to the assertions $A$.

\paragraph{Termination}
Truth of the verification conditions in a matrix program
only guarantees that no incorrect results can be produced.
It leaves open the possibility that no results appear
due to non-termination.
In my experience the development process is such that
one is aware of the reason it cannot give rise to an
infinite computation when a loop is closed.

\paragraph{Summary for the practitioner}
Hoare's verification method was widely seen as more important
than Floyd's. After all, the former is for {\sl structured}
programs, while Floyd's flowcharts reek of {\tt goto} statements.
Hoare acknowledged the inspiration from 
Floyd, while the public saw Hoare playing Jesus to
Floyd in the role of John the Baptist.
It is in the nature of things
who gets all the publicity.
But once one admits concern for correctness as the overriding one,
program structure may not offer any advantage.

But merely going back to flowcharts does not give anything in
exchange for sacrificing structure.
One would first face the difficulty of writing a correct flowchart 
only to be left with the difficulty of
finding assertions that yield true verification conditions.
It is only when one sees that a shortcut can be taken
that another path ahead is revealed.
The shortcut is to {\sl skip the flowchart} and to start writing
assertions on a blank slate, not encumbered by any flowchart,
with the specification of the programming problem as only constraint.
The first two assertions are the precondition
and the postcondition for the program as a whole.
Either a verification condition is found that connects the two
by feasible code or an intermediate assertion is found
that can be so connected. Thus the solution to the programming
problem is found by discovering missing assertions or
adding missing verification conditions.
All of this works because sets of verification
are already a programming language of sorts.
For the practitioner the technicalities of Section~\ref{sec:MC}
are dispensable: only this paragraph is needed, to be followed
by some of the examples.

\section*{{\large Acknowledgements}}
In writing this paper I am indebted to Paul McJones for
corrections and discussions.
In addition, the version of Egyptian multiplication presented
here is due to him.

\appendix
\section*{Appendixes}

\section{{\large The original Quicksort}}
\label{sec:origQS}
Here follows the transcription to C of Algorithms 63
and 64 \cite{hoareAlg63,hoareAlg64}.
\begin{verbatim}
void partition(int A[], int M, int N, int* i, int* j){
/* Makes X a random element of A[M..N] and permutes this
   segment in such a way that
     M <= J < I <= N provided M < N
     A[R] <= X for M <= R <= J
     A[R]  = X for J < R < I
     A[R] >= X for I <= R <= N
*/
  int I = *i; int J = *j; // using Hoare's I and J from now on
  int X; int F;
  F = rndm(M,N); X = A[F];
  I = M; J = N;
up: for(; I <= N; ++I)
      if (X < A[I]) goto down;
  I = N;
down: for(; J >= M; --J)
      if (X > A[J]) goto change;
  J = M;
change: if (I<J) { exchange(A, I, J);
                     ++I; --J;
                     goto up;
                 }
  else  if (I < F) {exchange(A, I, F); ++I;}
  else  if (F < J) {exchange(A, F, J); --J;}
  *i = I; *j = J;
}
void quicksort(int A[], int M, int N) { // sorts A[M..N]
  int I, J;
  if (M<N) { partition(A, M, N, &I, &J);
             quicksort(A, M, J);
             quicksort(A, I, N);
           }
}
\end{verbatim}

\section{{\large The original Quicksort verified}}
\label{app:dnfQS}
This is the transcription to C of the Liffig code in
Figure~\ref{prog:dnf} in
Section~\ref{sec:1961Part}.
Correctness has top priority.
Hence close similarity to the Liffig version is more important
than absence of code optimization opportunities.
\begin{verbatim}
void partition(int a[], int m, int n, int* i, int* j){
  int X; int F;
  F = rndm(m,n); X = a[F];
  int f=m; int s=f, t=n+1; goto A;
A: /* m<=k<f => a[k]=red &
      f<=k<s =>  a[k]=white
      s<=k<t => a[k] unknown
      t<=k<n+1 => a[k]=blue
      m<=f & f<=s & s<=t+1 & t<=n+1
   */
   if (s == t) { *j = f-1; *i = t; return; }
   if (s <  t) goto B;
   assert(0);
B: // A & s<t
   if (a[s] < X) {swap(a,f,s); ++f; ++s; goto A; }
   if (a[s] == X) { ++s; goto A; }
   if (a[s] > X) { --t; swap(a,s,t); goto A; }
   assert(0);
}
\end{verbatim}

\section{{\large The Quicksort of Foley and Hoare}}
Here follows the transcription to C of 
the Quicksort of Foley and Hoare \cite{foHoare71}.
\begin{verbatim}
void partition(int A[], int* ip, int* jp, int m, int n){
  int r, f; int i, j;
  f = (m+n)/2; r = A[f]; i = m; j = n;
  while (i <= j) {
    while (A[i]<r) ++i;
    while (A[j]>r) --j;
    if (i <= j) { swap(A, i, j); ++i; --j; }
  }
  *ip = i; *jp = j;
}
void Quicksort(int A[], int m, int n) {
  int i, j;
  if (m<n) {
    partition(A, &i, &j, m, n);
    Quicksort(A, m, j);
    Quicksort(A, i, n);
  }
}
\end{verbatim}

\section{{\large The Quicksort of Foley and Hoare: transcription
to C of the Liffig code}}
Here follows the transcription to C of 
the Liffig code for the Quicksort of Foley and Hoare
developed in Section~\ref{sec:FHpart}.
\begin{verbatim}
void partition(int a[], int* Ip, int* Jp, int m, int n){
  int r; int i,j;
S: r = (a[m]+a[n])/2 ; i = m; j = n; goto A;
H: *Ip = i; *Jp = j; return;
A: if (i == j) { goto F; }
   if (i  < j) { goto B; }
   assert(0);
B: if (a[i]  < r) { ++i; goto A; }
   if (a[i] >= r) { goto C; }
   assert(0);
C: if (a[j]  > r) { --j; goto D; }
   if (a[j] <= r) { swap(a,i,j); ++i; --j; goto E; }
   assert(0);
D: if (i == j) { goto F; }
   if (i  < j) { goto C; }
   assert(0);
E: if (i  > j) { goto H; }
   if (i <= j) { goto A; }
   assert(0);
F: if (a[i] <= r) { ++i; goto H; }
   if (a[j] >= r) { --j; goto H; }
   assert(0);
}
void Quicksort(int A[], int m, int n) {
  int i, j;
  if (m<n) {
    partition(A, &i, &j, m, n);
    Quicksort(A, m, j);
    Quicksort(A, i, n);
  }
}
\end{verbatim}

\bibliographystyle{plain}
\bibliography{bibfile}

\end{document}